\documentclass[11pt]{article}

\usepackage{amsmath}
\usepackage{amsthm}
\usepackage{amssymb}
\usepackage{hyperref}
\usepackage{enumitem}
\usepackage{color}
\usepackage{soul}
\usepackage{times}
\usepackage{graphicx}
\usepackage{wrapfig}
\usepackage{nicefrac}
\newcommand{\inv}[1]{\frac{1}{#1}}

\newcommand{\OT}{\nicefrac{1}{3}}
\newcommand{\TT}{\nicefrac{2}{3}}

\newcounter{smallitemizec}
\newenvironment{smallitemize}
{   \setcounter{smallitemizec}{0}
    \vspace{-2ex}
  \begin{list}{$\bullet$}
    {\usecounter{smallitemizec}
      \setlength{\parsep}{0pt}
      \setlength{\itemsep}{0pt}}
    }{ \end{list} 
   \vspace{-2ex}
}
\newcommand\E{\mathbb{E}}
\newcommand{\BA}{\begin{array}}
\newcommand{\EA}{\end{array}}
\newcommand{\BE}{\begin{enumerate}}
\newcommand{\EE}{\end{enumerate}}
\newcommand{\BI}{\begin{smallitemize}}
\newcommand{\EI}{\end{smallitemize}}

\newcommand{\I}{\item}

\def\BEQ#1\EEQ{\begin{align*}#1\end{align*}}
\def\BNEQ#1\ENEQ{\begin{align}#1\end{align}}

\newtheorem{claim}{Claim}
\newtheorem{lemma}{Lemma}
\newtheorem{theorem}{Theorem}
\theoremstyle{remark}

\newcommand{\BPF}{\begin{proof}}
\newcommand{\EPF}{\end{proof}}

\newcommand{\bracks}[1]{\left( {#1} \right)}
\newcommand{\stset}[2]{\left\{ {#1} \, \middle| \, {#2} \right\}}

\newcommand{\ifactor}{\textit{$\theta$}}
\newcommand{\speedup}{\textsf{\footnotesize speedup}}
\newcommand{\V}{\textsf{\footnotesize visited}}

\newcommand{\Ind}{\mbox{\sc non-coor}}
\newcommand{\Opt}{\mbox{\sc   coor}}

\usepackage{algorithm}
\usepackage{algpseudocode}

\begin{document}

\date{}
\author{Pierre Fraigniaud}
\author{Amos Korman}
\author{Yoav Rodeh}

\parskip0.1cm
\title{Parallel Exhaustive Search without Coordination}
\author{
Pierre Fraigniaud\thanks{CNRS and University Paris Diderot, France.  
 E-mail: {\tt pierre.fraigniaud@liafa.univ-paris-diderot.fr}. Additional support from the ANR project DISPLEXITY, and from the INRIA project GANG. }
 \and
 Amos Korman\thanks{CNRS and University Paris Diderot, France.  E-mail: {\tt amos.korman@liafa.univ-paris-diderot.fr}. This project has received funding from the European Research Council (ERC) under the European Union's Horizon 2020 research and innovation programme  (grant agreement No 648032). Additional support from the ANR project DISPLEXITY, and from the INRIA project GANG. 
}
\and
Yoav Rodeh
\thanks{Azrieli College of Engineering, Jerusalem, and The Weizmann Institute of Science, Israel. E-mail: {\tt  yoav.rodeh@gmail.com}.}
 }
\begin{titlepage}
\maketitle

\date{}

\def\thefootnote{\fnsymbol{footnote}}

\begin{abstract}

We analyze parallel algorithms in the context of \emph{exhaustive search} over totally ordered sets. Imagine an infinite list of ``boxes'', with a ``treasure'' hidden in one of them, where the boxes' order reflects the importance of finding the treasure in a given box. At each time step, a search protocol executed by a searcher has the ability to peek into one box, and see whether the treasure is present or not. Clearly, the best strategy of a single searcher would be to open the boxes one by one, in increasing order. Moreover, by equally dividing the workload between them, $k$~searchers can trivially find the treasure $k$ times faster than one searcher. However, this straightforward strategy is very sensitive to failures ({\em e.g.,} crashes of processors), and overcoming this issue seems to require a large amount of communication. 
We therefore address the question of designing parallel search algorithms maximizing their {\em speed-up} and maintaining high levels of {\em robustness}, while minimizing the amount of resources for coordination. 
Based on the observation that algorithms that avoid communication are inherently robust, we focus our attention on identifying the best running time performance of {\em non-coordinating} algorithms. Specifically, we devise non-coordinating algorithms that achieve a speed-up of $9/8$ for two searchers, a speed-up of $4/3$ for three searchers, and in general, a speed-up of $\frac{k}{4}(1+1/k)^2$ for any $k\geq 1$ searchers. 
Thus, asymptotically, the speed-up is only four times worse compared to the case of full-coordination. 
Moreover, these bounds are tight in a strong sense as no non-coordinating search algorithm can achieve better speed-ups.
Furthermore, our algorithms are surprisingly simple and hence applicable.
Overall, we highlight that, in faulty contexts in which coordination between the searchers is technically difficult to implement, intrusive with respect to privacy, and/or costly in term of resources,  it might well be worth giving up on coordination, and simply run our  non-coordinating exhaustive search algorithms.

\end{abstract}

\thispagestyle{empty}
\end{titlepage}

\setcounter{page}{1}

\section{Introduction}

BOINC \cite{BOINC}  (for Berkeley Open Infrastructure for Network Computing) is a platform for volunteer computing supporting dozens of projects such as the famous SETI@home analyzing radio signals for identifying signs of extra terrestrial intelligence. Most projects maintained at BOINC use parallel exhaustive search mechanisms where a central server controls and distributes the work to volunteers (who process this work during unused CPU and GPU cycles on their computers). The framework in this paper is a potential abstraction for projects operated at platforms similar to BOINC aiming at tackling exhaustive search in a totally ordered set like, {\em e.g.,} breaking encryption systems with variable key length, with hundreds of thousands searchers. 

In general, parallel algorithms \cite{JaJa,Tom} are algorithms that are concurrently executed on potentially  many different processing devices. Such algorithms are often evaluated with respect to their {\em speed-up}, that is, how much faster the parallel algorithm with $k$ processors runs in comparison to the best running time that a single processor can achieve.  To obtain a large speed-up, the algorithm should  typically enable the processors to coordinate their operations for balancing the work load between them as evenly as possible. Such a coordination effort however comes at a cost, as it often requires additional computation and/or communication steps. In fact, in some cases, the additional overhead involved in coordinating the processors can overshadow the speed-up altogether, and result in a {\em slowdown} of the parallel computation if too many processors are used. In parallel computing, there is therefore an inherent tradeoff, which depends on the targeted problem, between the amount of coordination required to solve the problem efficiently, and the cost of this coordination. 

One extremity of the spectrum is the class of {\em non-coordinating} algorithms, which are parallel algorithms whose computing entities are operating independently with no coordination. In such an algorithm, all processors execute the same protocol, differing only in the outcome of the flips of their random coins
, like, {\em e.g.,} searching in a graph using parallel random walks~\cite{AlonZvika}. Most problems cannot be efficiently parallelized without coordination. However, when such parallelization can be achieved,  the benefit can potentially be high not only in terms of saving in communication and overhead in computation, but also in terms of robustness.

To get an intuition on why non-coordinating algorithms are inherently \emph{fault-tolerant}, let us focus on parallel {\em search} problems where the goal is to minimize the time until one of the searchers finds a desired object. When executing a non-coordinating algorithm in such  contexts, the correct operation as well as the running time performances can only improve  if more processors than planned are actually being used. Suppose now that an oblivious adversary is allowed to {\em crash} at most $k'$ out of the $k$ processors at certain points in time during the execution. 
To overcome the presence of $k'$ faults, one can simply run the algorithm that is designed for the case of $k-k'$ processors. If the speed-up of the algorithm without crashes is $\speedup(k)$, then the speed-up of the new {\em robust} algorithm would be $\speedup(k-k')$. Note that even with coordination, one cannot expect to obtain robustness at a cheaper price since the number of processors that remain alive is in the worst case $k-k'$. This strong robustness property of parallel non-coordinating algorithms motivates the study of this class of algorithms, and in particular the investigation of the best possible speed-up that they can  achieve. 

We propose to formally evaluate the impact of having no coordination using a competitive analysis approach by introducing the notion of \emph{non-coordination ratio}. 
This notion is meant to compare the best possible performances of a parallel algorithm in which the computing entities are able to fully coordinate, with the best possible performances of a parallel algorithm whose computing entities are operating independently. More 
formally, for a given problem,  let us denote by $\speedup_{\scriptsize \Opt}(k)$  the largest speed-up in expected running time  that can be achieved 
 when coordination between the processors comes at no cost. 
Similarly, let $\speedup_{\scriptsize \Ind}(k)$ denote  the largest speed-up in expected running time that can be achieved by a non-coordinating algorithm with $k$ processors. The {\em non-coordination ratio} for the problem  with $k$ processors is then defined as:
 \[
 \rho(k) ~=~ \frac{\speedup_{\scriptsize\Ind}(k)}{\speedup_{\scriptsize\Opt}(k)}~.
 \]
Note that $0< \rho(k)\leq 1$ for every integer~$k\geq 1$. A non-coordination ratio close to~1 indicates that the problem is essentially oblivious to coordination and a ratio close to~$0$ indicates that  the problem presents a high sensitivity to coordination. 
%

One class of fundamental problems that may enjoy a large non-coordination ratio in some circumstances, is the class of search problems over totally ordered sets,  which are often tackled using \emph{exhaustive search} algorithms.  The objective of such \emph{linear search} problems is to find a solution among a set of candidate solutions that are linearly ordered according to their quality. For instance, searching for a proper divisor of a random number $n$ is an illustration of linear search. Indeed, enumerating the candidate divisors in increasing order, from 2 to $n - 1$, and checking them one by one, is the typical approach to solve the problem, since the probability that $n$ is divisible by a given prime is inversely proportional to this prime. Similarly, in cryptography, an exhaustive search attack is better proceeded by systematically checking smaller keys than longer ones, as the time to check a key is typically exponential in its size. In general, linear search appears in contexts in which the search space can be ordered in a way such that, given that the previous trials were not successful, the next candidate according to the order is either the most preferable, or most likely to be valid, or the easiest to check.

In this paper, we focus on one basic linear search problem, called, the {\em treasure-hunt} problem.
Formally, consider an infinite ordered list of ``boxes'', $(B_1,B_2,\ldots)$, with an adversary that hides a ``treasure'' in one of the boxes, $B_x$, for some index~$x$. The boxes are listed in an order that reflects the importance of finding the treasure in a given box. That is, finding the treasure hidden in $B_i$ for small values of~$i$ is more urgent than for large values of~$i$. A \emph{search protocol} is unaware of the index~$x$ of the box where the treasure has been placed, and is aiming at finding that treasure as fast as possible. Time proceed in discrete steps. At each time step, a protocol executed by a searcher has the ability to peek into one box and see whether the treasure is present or not in this box. The protocol is  terminated once one of the searchers finds the treasure.

 In the case of a solo searcher, the exhaustive search  algorithm will find the treasure in $x$ time, and no solo searcher can perform faster than $x$. Hence, for a given algorithm $A$ running with $k$ searchers, we  measure the {\em speed-up} function of $k$ searchers with respect to $x$ as:
\[
\speedup_{A}(k,x)=\frac{x}{\E(\text{time to find $x$ with $k$ searchers running  $A$})}
\]
We define the speed-up of Algorithm~$A$ with respect to $k$ searchers as:
\[
\speedup_{A}(k)= \liminf_{x \to +\infty}\speedup_{A}(k,x)~.
\] 
The exhaustive search  strategy of a solo searcher can be trivially parallelized to yield a speed-up of $k$. 
For this, the $k$ searchers simply need to enumerate themselves from~1 to~$k$, and searcher  $1\leq i \leq k$,~peeks~into~box $B_{i+(t-1)k}$ at time $t\geq 1$. Note however, that this algorithm is highly sensitive to faults. For example, without additional operations, the crash of a single searcher at any time during the execution can already reduce the speed-up to zero!  In addition, even if none of the searchers crashes, the running~time of this algorithm highly depends on the assumption that all searchers agree on the precise ordering $(B_1,B_2,\ldots)$ of the boxes, and, in fact, the algorithm can have very poor performances in case of even tiny fluctuations in the ordering of input boxes by the searchers\footnote{For example, with two searchers, 
consider the case where the first searcher orders the boxes correctly as $(B_1,B_2,\ldots)$, but the second searcher mistakenly adds an additional box $B'$ between $B_i$ and $B_{i+1}$, for some integer $i$.
Then there are infinitely many instances for which none of the two searchers will ever find the treasure. 
A similar example can occur when the set of input boxes is the same for both searchers, but their relative orders are slightly different.}. The implementation of the trivial exhaustive search  algorithm in such cases may  require a significant amount of coordination and communication between the searchers. 
 These examples indicate that to achieve robustness, more complex algorithms should be considered. 

 As mentioned, a class of parallel algorithms that are natural to consider in fault-tolerant settings is the class of non-coordinating algorithms.  
We ask the following  questions regarding the performances of such algorithms:
First, phrased in terms of two searchers, can  two non-coordinating searchers speed up the search by a factor close to two?  More generally, what is the best possible speed-up that can be achieved by $k$ non-coordinating searchers, or, in other words, what is the non-coordination ratio $\rho(k)={\speedup_{\scriptsize\Ind}(k)}/{k}$?  Note that it is not  clear at a first glance whether the non-coordination ratio remains bounded from below by a positive constant as $k$ goes to infinity. 

\subsection{Our results}

In a nutshell, we entirely solve the issue of non-coordination in the treasure-hunt problem by precisely identifying the non-coordination ratio for any number of searchers. Specifically, we prove that the  non-coordination ratio of $k$ searchers is:
\[
\rho(k)=\inv{4}\left(1+\inv{k}\right)^2~.
\]
This means that the best speed-up that can be achieved is $\frac98$ for two searchers, $\frac43$ for three searchers, and roughly $\frac{k}{4}$ for $k$ searchers, as $k$ grows larger. 


Interestingly, the non-coordinating algorithm achieving the aforementioned  ratio  for $k$ searchers~is~so~simple that it can described in just a few lines. 
Define the sequence $I_i=\{1,2,\ldots, i(k+1)\}, \; i\geq 1$, of  nested  sets. 
The algorithm first picks a box whose index is chosen uniformly at random from $I_1$. Then, it picks another index in $I_1$, avoiding the  already choosen index. In rounds 3 and 4, the algorithm chooses two indices in $I_2$, according to the same principle, and so forth. 
Formally, the algorithm is described as follows (note that for each $k$ the algorithm is different).

\begin{algorithm}[htb]
\begin{algorithmic}[1]
\State{Let $\V_s(t)$ denote the  set of indices indicating boxes visited by $s$ before time $t$.}
\State{At time $t$, peek into $B_i$ where $i$ is chosen u.a.r. from $I_{\lceil t/2\rceil}\setminus\V_s(t)$.}
\end{algorithmic}
\caption{non-coordinative search with $k$ searchers: program of an arbitrary searcher $s$}
\label{searchalgo}
\end{algorithm}

As mentioned earlier, since we are dealing with non-coordination search algorithms, we immediately get {\em robustness} with respect to {\em crashes} of searchers. 
In addition, our upper bound holds even in the case where the searchers do not have the same ordering of boxes or even the same list of boxes. 
In order to state this robustness property formally, we introduce the following notions. Let $L=(B_1,B_2,\dots)$ denote the correct sequence of boxes.  We assume that all boxes in $L$ appear in the list of each searcher, but that searchers may have other boxes as well. Moreover, in the ``eyes'' of a given searcher, the boxes in $L$ may not appear in the same relative ordering as in $L$. For every $i$, mark by $\sigma_s(i)$ the box index of $B_i\in L$ in the ``eyes'' of searcher~$s$. We show that, as long as for each searcher $s$, it is guaranteed that $\lim_{i \rightarrow \infty} \sigma_s(i)/i=1$,  then the speed-up remains precisely the same. 

To analyze treasure-hunt algorithms, we first observe that the crucial aspects of non-coordinating algorithms can be represented by infinite matrices. This interpretation readily allows us to prove the required robustness property for all non-coordinating algorithms, as well as a lower bound proof for the case of a solo-searcher.  We then turn our attention to prove that the speed-up of Algorithm~\ref{searchalgo} is, for every $k$, at least ${(k+1)^2}/{4k}$. This proof relies on carefully analyzing the corresponding matrix and on using known properties of the Gamma function.
Finally, we prove that no algorithm has a speed-up better than 
${(k+1)^2}/{4k}$.  This proof is technically the hardest. To establish it, we first approximate matrices by continuous functions. We then turn to examine a weighted average of the inverse of the speed-up, where the average is taken over the indices larger than some fixed index. By showing a lower bound on that we actually show an upper-bound on the speed-up of all algorithms. Choosing the weighted average carefully we arrive at our result. The two complementing bounds yield the exact value of $ \inv{4}\bracks{1 + \frac{1}{k}}^2$ for the non-coordination ratio $\rho(k)$.

To sum up, our upper bound on the non-coordination ratio implies  that there is an incompressible price to be paid for the absence of coordination, which is asymptotically a factor of four away from an ideal optimal algorithm which performs with perfect coordination, but whose coordination costs are not accounted for. This price is incompressible in the sense that no non-coordinating algorithms can do better than that. On the other hand, this price is actually  reasonably low, and, again, it is computed by competing against an ideal optimal algorithm, where coordination costs are ignored. Therefore, in faulty contexts in which coordination between the searchers may yield severe overheads  ({\em e.g.,} when the searchers are subject to ordering errors, and/or when searchers can crash), it might well be worth giving up on coordination, and simply run our non-coordinating algorithm.

\subsection{Related work}



The treasure-hunt problem on the line was first studied in the context of the \emph{cow-path} problem \cite{BCR91}, which became classical in the framework of online computing. 
One major difference between that problem and the setting we consider, is that in the cow-path problem, in order to visit a new location (box in our terminology), the agent cannot simply ``jump'' to that location, and instead, it must visit all intermediate locations. 
More specifically, that problem assumes a mobile searcher that is initially placed at the zero point, and an adversary that hides a treasure at point $x$ (either negative or positive). The searcher is unaware of the value $x$, and even of $x$'s sign, and its objective is to find the treasure as quickly as possible. 
To move from point $i$ to point~$j$ the searcher needs to pass through all intermediate points, and hence pays a cost of $|i-j|$ for such a move. Therefore, in any case, the searcher must pay a total cost of $d=|x|$ merely to travel to the treasure position. It was established in~\cite{BCR91} that the best performances that any deterministic algorithm can achieve is $9d$. The algorithm that achieves this performance follows a straightforward exponential search approach. The randomized version of this problem was studied \cite{KRT96}, showing that the best expected time for finding the treasure is roughly half of what a deterministic algorithm can achieve. Note that in any case, no matter how many agents are employed, how they coordinate, and whether they are deterministic or randomized, if all of them are initialized at the zero point of the line, a cost of $d$ could not be avoided. This means that the cow-path problem on the line cannot be effectively parallelized. 
 
Variants on the cow-path problem were also studied on multiple intersecting lines, on the grid, and on trees. In particular, it was  shown in~\cite{BCR91} that the \emph{spiral search} algorithm is optimal in the two-dimensional grid, up to lower order terms.  
Motivated by applications to central search foraging by desert ants, the authors in \cite{FK12,FKLS} considered the ANTS problem, a variant of the cow-path problem on the grid, and showed that a speed-up of $O(k)$ can be achieved, with $k$ independent searchers. Emek et al. showed in \cite{Uitto} that the same asymptotic speed-up can be achieved even with $k$ searchers that use bounded memory as long as they are allowed to communicate. Several other variants of the cow-path problem and the ANTS problem where studied in  \cite{DFG06,Emek-ants,KSW86,Tobias,Nancy-PODC}.


In a series of papers on parallel random walks in graphs,  a speed-up of $\Omega(k)$ is established for various finite graph families, including, {\em e.g.,} expanders and random graphs \cite{AlonZvika, ES11, CFR09}.


%
%
%
%

\subsection{Terminology}
Our universe contains boxes indexed by $i \in \{1,2,...\}$, and an adversary places a treasure in one of them. At each time step, a searcher can peek into exactly one box. There are $k$ searchers and they are all completely identical in that they have the same algorithm, yet their randomness is independent. Until one of them finds the treasure, they cannot communicate at all. The aim of the searchers is to maximize the speed-up by minimizing the expected time until one of them finds the treasure. In our technical discussion it will be often easier to work with the {\em inverse} of the speed-up. Specifically, let us define:
\[
\ifactor(k,x) = \frac{\E(\text{time to find $x$ with $k$ searchers})}{x} \mbox{~~~~and~~~~~} \ifactor(k) = \limsup_{x \rightarrow \infty} (\ifactor(k,x))
\]
So, an algorithm with $\ifactor(3) = 1/2$ means that running this algorithm on three searchers will result in an expected running time that is twice as fast as the trivial one-searcher algorithm.

\section{From Algorithms to Matrices}

Our first step in analyzing different (non-coordinating) algorithms is to consider the following infinite matrix, where we 
at first think of just one searcher.
Given an algorithm, write down a matrix $N$, where
$N(x,t)$ marks the probability that the algorithm has not visited box $x$ up to (and including) step $t$. 


\begin{wrapfigure}{r}{0.47\textwidth}
\vspace{-25pt}
\BEQ
\BA{c|ccccccc}
_{x\downarrow}^{t\rightarrow} & 0 & 1 & 2 & 3 & 4 & 5 & 6 \\
\hline
1 & 1  & \TT  & \OT & 0    & 0    & 0  & 0  \\
2 & 1  & \TT  & \OT & 0    & 0    & 0  & 0  \\
3 & 1  & \TT  & \OT & 0    & 0    & 0  & 0  \\
4 & 1  & 1    & 1   & 1    & \TT  & \OT  & 0  \\	
5 & 1  & 1    & 1   & 1    & \TT  & \OT  & 0  \\
6 & 1  & 1    & 1   & 1    & \TT  & \OT  & 0  \\
\EA
\EEQ
\vspace{-30pt}
\end{wrapfigure}
\smallskip
An example:
We look at the following algorithm. It chooses a box to peek into at random from the boxes $0,1,2$. Then again one of the two that was not looked into and then the third. Then it moves to consider the boxes $3,4,5$, and so on. On the right hand side, we see how the matrix $N(x,t)$ for this algorithm starts.

\bigskip
\noindent Some observations:
\smallskip
\BI
\I
Each row is monotonically non-increasing. If we wish the algorithm to have a bounded expected time for all $x$'s, then the limit of each row has to be 0 (but this is only a necessary condition).
\I
The sum of row $x$ is the expected time until the algorithm peeks into box $x$.
Indeed, let $I_{x,t}$ denote the indicator random variable that is 1 iff $t <$ the visit time of $x$.
The sum of these over $t$ is the visit time. Also, 
$\Pr[I_{x,t} = 1] = N(x,t)$,
so we get the result by linearity of expectation. This means:
\BEQ
\ifactor(1,x) = \inv{x} \sum_{t=0}^\infty N(x,t)
\EEQ
\I
Given the matrix $N$ for one searcher, what would be the $N$ matrix for $k$ searchers? The probability of~$x$ not being looked into up to step $t$ is the probability that all $k$ searchers didn't peek into it, which is $N(x,t)^k$. So by the same reasoning as the last point, we get:
\BNEQ\label{eq:theta}
\ifactor(k,x) = \inv{x} \sum_{t=0}^{\infty} N(x,t)^k 
\ENEQ
\parskip-0.1cm
\I
Since $1-N(x,t)$ is the probability that box $x$ was peeked into by step $t$, summing these numbers over column $t$, we get the expected number of boxes checked by this time,
which is of course at most~$t$.
\EI

\bigskip
\noindent In an algorithm that remembers what boxes it already looked into and takes care not to redundantly check the same box twice, this last point becomes an equality. We can then write that for all $t$:
\BNEQ \label{eq:columns}
\sum_x \bracks{1 - N(x,t)} = t
\ENEQ 
\parskip0.1cm
Indeed, if we have an algorithm that does not behave this way, we alter it as follows. Run it as usual but remember every box checked. Then, every time the algorithm wants to peek into a box it already checked, it instead looks into some other box $x$ that was not visited yet\footnote{There is a somewhat delicate point here. For example, the given algorithm could react badly when peeking into a box it did not expect. Therefore, in the modified algorithm, when we check a new box, we do not ``tell" the algorithm that we did this. Meaning that in all the internal decisions of the algorithm it will behave as the original algorithm.}. The new algorithm can only improve on the original in terms of speed-up, since its $N$ matrix will have smaller values. From now on we shall assume that \eqref{eq:columns} always applies.

\subsection{Lower Bound For One Searcher}
The trivial exhaustive search with a solo-searcher achieves a speed-up of 1. 
Since the definition of speed-up concerns the {\em asymptotic}  behavior of algorithms, it is a-priori not  clear that 
 a single randomized searcher cannot do better than  exhaustively search the boxes. The following theorem states that there is no surprise here, and indeed exhaustive search is the best strategy.\medskip
\begin{theorem}\label{thm:lower-solo}
Any algorithm has $\ifactor(1)\geq 1$.
\end{theorem}
\medskip

\noindent The proof is quite simple, yet we show it for completeness and as an example for using the matrix representation for proving lower bounds. It also illustrates the basic proof technique we will use for the more difficult case of $k \geq 2$: Since any one particular box $x$ can have a good $\ifactor(k,x)$, we take the average of many $\ifactor(k,x)$ and show a lower bound on that.
\BPF
Take some algorithm that has $\ifactor(1) = \alpha$. This means that for any $\epsilon > 0$ there is a box $s$, such that for all  $x \geq s$, we have
$
\sum_{t=0}^\infty N(x,t) \leq (\alpha + \epsilon) x
$.
Take some large $M$, and sum all  rows from $s$ to  $M$:
\BEQ
\sum_{x=s}^M \sum_{t=0}^\infty N(x,t) 
&\leq \sum_{x=s}^M (\alpha + \epsilon) x 
 = (\alpha + \epsilon) \frac{(M-s+1)(M+s)}{2} 
\EEQ
On the other hand, we have:
$
t = \sum_{x=0}^\infty (1 - N(x,t)) 
\geq \sum_{x=s}^M (1 - N(x,t)) = 
M - s +1- \sum_{x=s}^M N(x,t), 
$
for each $t$.  Therefore:
\BEQ
\sum_{x=s}^M N(x,t) \geq M -s+1 - t
\EEQ
Now again, look at the double sum:
\BEQ
\sum_{x=s}^M \sum_{t=0}^\infty N(x,t)
\geq  \sum_{x=s}^M \sum_{t=0}^M N(x,t)
= \sum_{t=0}^M \sum_{x=s}^M N(x,t)
 \geq  \sum_{t=0}^M (M-s+1-t)  =\frac{(M+1)(M-2s+2)}{2}
\EEQ
Combining, we get that:
\BEQ
(\alpha + \epsilon) \frac{(M-s+1)(M+s)}{2} &\geq \frac{(M+1)(M-2s+2)}{2}
\EEQ
If $\alpha < 1$, take a sufficiently small $\epsilon$ so that $\alpha + \epsilon < 1$.
Now, taking $M$ to infinity, both fractions behave asymptotically like~$M^2/2$, and therefore  the inequality is not satisfied. It follows that  $\alpha \geq 1$.
\EPF

\subsection{Non-Coordination is Robust}
We have already seen that non-coordinating search algorithms are highly robust with respect to crashes of searchers. Here we look at the case that searchers do not crash, but each searcher may hold a different view of the numbering of boxes. Even a small difference in these ordering may be devastating to some algorithms, and yet, we show that in the case of non-coordinating processes, 
this has actually little affect, as long as the numbers are not way off. The proof of the following theorem is deferred to Appendix \ref{apx:robust}. It is simple and relies on a generalized form of the Cauchy-Schwartz inequality.
\medskip 
\begin{theorem}\label{thm:robust}
Denote $c = \speedup(k)$ when all searchers see the correct order of the boxes. Consider the case where they see a different ordering (possibly including some extra boxes), and mark by $\sigma_s(i)$ the index of box $i$ in the eyes of searcher~$s$. 
If for every $i$, $\lim_{i \rightarrow \infty} \sigma_s(i)/i=1$
then the new speed-up is at least~$c$.
\end{theorem}

\section{The Speed-Up of Algorithm \ref{searchalgo} is at least ${(k+1)^2}/{4k}$} \label{sec:upper}
The following theorem states that for every integer $k$, the non-coordination ratio is $\rho(k) \geq \inv{4}\bracks{1 + \frac{1}{k}}^2$.
\medskip

\begin{theorem}
The speed-up of Algorithm \ref{searchalgo} is at least ${(k+1)^2}/{4k}$.
\end{theorem}
\medskip
\begin{wrapfigure}{r}{0.3\textwidth}
\vspace{-25pt}
\BEQ
\BA{c|ccccccc}
   & 0 & 1 & 2 & 3 & 4  \\
\hline
1 & 1 & \TT  & \OT & \nicefrac{1}{4}     & \nicefrac{1}{6} \\
2 & 1 &  \TT  & \OT & \nicefrac{1}{4}     & \nicefrac{1}{6} \\
3 & 1 & \TT  & \OT & \nicefrac{1}{4}     & \nicefrac{1}{6} \\
4 & 1 & 1    & 1   & \nicefrac{3}{4} & \nicefrac{1}{2} \\
5 & 1 & 1    & 1   & \nicefrac{3}{4} & \nicefrac{1}{2} \\
6 & 1 & 1    & 1   & \nicefrac{3}{4} & \nicefrac{1}{2} \\
\EA
\EEQ
\vspace{-20pt}
\end{wrapfigure}

\noindent Recall that a searcher operating under Algorithm \ref{searchalgo} first peeks into a box with index chosen uniformly in the set $\{1,\ldots,k+1\}$. It then chooses 
another index in that domain omitting the one already chosen. Subsequently, it chooses an index uniformly in $\{1,\ldots,2(k+1)\}$, omitting the two that were already chosen. Then a fourth one in the same domain, etc. 
It is convenient to inspect the algorithm in its matrix form. 
For example, for $k=2$, the $N(x,t)$ matrix is illustrated on the right.

\smallskip
\noindent {\em Proof.}
Ignoring the first column (for large $x$ its contribution will be negligible), we partition the matrix to blocks of size $(k+1) \times 2$ each, where the rows of each block are of equal values.
Let us start by ignoring the odd columns, and so we focus only on the bottom right corner of each block.
For integers $x$ and $t$, mark $b(x,t) = N((k+1)x, 2t)$.  
Note that for $x > t$ we have $b(x,t) = 1$. 
On the other hand, for $t \geq x$, since we are randomly choosing two different indices out of $t(k+1) - 2(t-1) = t(k-1) + 2$
still unchosen indices, we have:
\[
b(x,t) 
= b(x,t-1) \cdot \left(1 - \frac{2}{t(k-1) + 2}\right)
= b(x,t-1) \cdot \frac{t}{t + 2/(k-1)}
\]
Mark $\delta = 2/(k-1)$, and we get:
\BEQ
b(x,t) = 
\frac{t}{t + \delta} \cdot
\frac{t-1}{t-1 + \delta} \cdot ... \cdot
\frac{x}{x + \delta} = \prod_{i=x}^t \frac{i}{i + \delta}
\EEQ
We return to $\ifactor(k,x)$:
\BEQ
\ifactor(k,x) 
 & = \inv{x} \sum_{t=1}^\infty N(x,t)^k 
\leq \inv{x} \left( 1 + 2\sum_{t=1}^\infty N(x,2t)^k \right)	= 
\inv{x} + \frac{2}{x} \sum_{t=1}^\infty b\bracks{\left\lceil{x}/{(k+1)}\right\rceil,t}^k
\EEQ
Where the inequality is because $N$ is monotonically decreasing and so $N(x,2t+1) \leq N(x,2t)$. This accounts for all odd $t$'s except $t=1$, and that is why we add the $+1$.  	
Mark $x' = \left\lceil{x}/{(k+1)}\right\rceil$. Recalling that for $t<x$, we have $b(x,t) = 1$, we obtain the following:
\BNEQ \label{eq:upper1}
\ifactor(k,x) \leq \inv{x} + \frac{2}{x} \left(
x' +
\sum_{t=x'}^\infty \bracks{\prod_{i=x'}^t \frac{i}{i + \delta}}^k
\right)
= 
\inv{x} + \frac{2x'}{x}\left( 1 +  
\inv{x'}\sum_{t=x'}^\infty \bracks{\prod_{i=x'}^t \frac{i}{i + \delta}}^k
\right)
\ENEQ
Using several properties of the Gamma function we prove the following claim in Appendix \ref{apx:upper}:
\medskip
\begin{claim}\label{claim:boxes}
For any $\delta > 0 $ and $k > 2/\delta$, 
we have: $\limsup_{x\rightarrow\infty}\inv{x} \sum_{t=x}^\infty \bracks{\prod_{i=x}^t  \frac{i}{i + \delta}}^{k} \leq \inv{\delta k - 1}~.$
\end{claim}
\smallskip
\noindent Since $\ifactor(k) = \limsup_{x\rightarrow\infty} \ifactor(k,x)$, then plugging Claim \ref{claim:boxes}   in Equation  \eqref{eq:upper1}, and taking $x$ to infinity, we get:
\BEQ
\ifactor(k)\leq 0 + \frac{2}{k+1}\bracks{1 +  \inv{\delta k - 1}}
= 
\frac{2}{k+1} \cdot \bracks{1 +  \inv{{2k}/{(k-1)} - 1}}
= 
\frac{2}{k+1} \cdot \bracks{1 + \frac{k-1}{k+1}}
= \frac{4k}{(k+1)^2}
\EEQ
The theorem now follows as $\ifactor(k)$ is the inverse of the speed-up. 
\qed

\section{The Best Possible Speed-Up is Precisely ${(k+1)^2}/{4k}$}
We show that Algorithm \ref{searchalgo} is in fact optimal for every $k\geq 1$.
This also shows a matching lower bound for the non-coordination ratio of the treasure-hunt problem, and so $\rho(k) = \inv{4}\bracks{1 + \frac{1}{k}}^2$.
\medskip
\begin{theorem} \label{thm:lowerBound}
Any algorithm for $k$ searchers has $\ifactor(k) \geq {4k}/{(k+1)^2}$.
\end{theorem}
\smallskip

\noindent In Theorem \ref{thm:lower-solo} we have seen that the statement in Theorem \ref{thm:lowerBound} holds for the case $k=1$. The remaining of this section is dedicated to analyzing the cases where $k\geq 2$.
Our strategy for proving that  $\ifactor(k)= \limsup_{x\rightarrow \infty} \ifactor(k,x)$ is at least ${4k}/{(k+1)^2}$ is by showing that for every $s$ there is some $x > s$ such that $\ifactor(k,x)$ is greater than this value. Hence, in what follows, we fix an integer $s$. 

We will show that for any algorithm, there is some weighted average of the values $\ifactor(k,x)$ for the $x$'s that are greater than $s$, such that this average is at least 
${4k}/{(k+1)^2}$. This means there is at least one $x \geq s$ for which  $\ifactor(k,x) \geq {4k}/{(k+1)^2}$, proving the result.
Informally, the reason we take a weighted average and not deal with the $\limsup$ directly, is that it is easier to work with sums of sums  than with $\limsup$ of sums.

For our lower bound we turn to the continuous setting, 
and view the matrix $N$ as a continuous function:
\BEQ
N : [s, \infty) \times [0,\infty) \rightarrow [0,1]
\EEQ
Our equations will be stated as integrals instead of sums.	
Proving a lower bound on the continuous version easily translates to a lower bound on the discrete version, since we can approximate a step-function as closely as we wish with continuous functions, and all integrals will approximate sums to any precision wanted.

Note that we are ignoring the behavior of $N$ on all $x<s$, since we do not care about their speed-up. 
An optimal algorithm for this case (with a fixed $s$) will not even try to peek into an $x < s$ (in a similar way to the argument below  \eqref{eq:columns}). We can therefore write 
our column requirement of  \eqref{eq:columns} as follows: For every~$t$,
\BNEQ \label{eq:intRestriction}
t = \int_s^\infty 1 - N(x,t) dx
\ENEQ
Similarly, the continuous equivalent of Equation \eqref{eq:theta} is $\ifactor(k,x)=\inv{x}{\int_0^\infty  N(x,t)^k dt}$.

Mark by $\omega(x)$ the weight we give $\ifactor(k,x)$ in our weighted average.
This means that $\int_s^\infty \omega(x) dx= 1$ and $\omega(x)$ is always between 0 and 1. The value that we wish to bound from below is:
\BNEQ \label{eq:targetInt}
\int_s^\infty \omega(x) \cdot \ifactor(k,x) dx =  
\int_s^\infty \omega(x)\frac{\int_0^\infty  N(x,t)^k dt}{x} dx
=\int_0^\infty \int_s^\infty  \mu(x) N(x,t)^k dx dt
\ENEQ
where we mark $\mu(x) = \omega(x)/x$, 
and exchange the order of integrals. This is fine by Tonelli's theorem \cite{Tonelli}, as the function we integrate is continuous and always non-negative.
We will try to analyze as much as possible with a general $\mu$, and later plug in a specific weight function $\omega$ to get our result.

We assume that $\mu$ is strictly positive,  bounded, continuous
and  monotonically decreasing.
We note that $\int_s^\infty \mu(x) dx$ is defined, since $\mu$ is continuous and its integral is bounded:
\BNEQ \label{eq:muIntegrable}
\int_s^\infty \mu(x) dx = \int_s^\infty \frac{\omega(x)}{x} dx \leq 
\frac{1}{s} \int_s^\infty \omega(x) dx = \frac{1}{s}
\ENEQ

\subsection{An Optimal Function $N$}

We next proceed to show, that given a specific weighted average $\omega$ (and thus a specific $\mu$), we can find one quite simple $N$ that minimizes our target integral \eqref{eq:targetInt}. The value of the integral on this $N$ would then be our lower bound.

Under the restrictions that $N(x,t)\in [0,1]$ and  $\int_s^\infty (1- N(x,t))dx = t$, 
we wish to find a function $N$ that minimizes the corresponding inner integral of \eqref{eq:targetInt},
{\em i.e.,} $\int_s^\infty \mu(x) N(x,t)^k dx$. Observe that for different values of $t$, 
 all these integrals are  independent of each other. Therefore, we can look at each one separately and find the optimal $N(x,t)$ for each specific~$t$. Intuitively, for a fixed $t$, we will make use of the fact that we can always move  a small amount of  ``mass" from one $N(x,t)>0$ to another $N(x',t)<1$, without violating the restrictions. Finding the optimal balance to optimize the target integral is the idea behind the proof of the following lemma.
\bigskip
\begin{lemma} \label{lm:balance}	
Fix $s\geq 0$ and some $M > 0$. For continuous functions $a: [s,\infty) \rightarrow (0,M]$ and $f : [s,\infty) \rightarrow [0,1]$ where $\int_s^\infty a(x) dx = A$ and $\int_s^\infty (1-f(x))dx = T$, 
the minimum of $\int_s^\infty a(x) f(x)^k dx$, ranging over all possible $f$'s is achieved when
$f(x) = \min(1, \alpha / a(x)^{1/(k-1)})$, where $\alpha$ is a function of $a$ and $T$, and independent~of~$x$. 
Also, fixing~$a$, $\alpha$ is a continuous function of $T$.
\end{lemma}
\medskip
\noindent We prove Lemma \ref{lm:balance} in Appendix \ref{apx:optimal}. The proof actually requires more than the simple proof strategy of re-balancing of masses, as the space of possible solutions is not compact.
Using this lemma,  an optimal~$N$~is:
\BNEQ \label{eq:N}
N(x, t) = \min(1, \alpha / \mu(x)^\inv{k-1})
\ENEQ
where $\alpha$ is a continuous function of $t$ (and yet, for readability, we don't write $\alpha(t)$). This of course extends to the double integral \eqref{eq:targetInt}, since no other $N$ can get a smaller value for $\int_0^\infty \int_s^\infty \mu(x) N(x,t)^k dx dt$, as our $N$ would beat it in every inner integral. From this point onwards, $N$ will refer to this specific function.

Note that $N$ is continuous, since both $\mu$ is continuous as a function of $x$ and $\alpha$ is continuous~as~a~function~of~$t$.
We would like to calculate our double integral on this particular $N$ which would give us a lower bound on the weighted average $\omega$ of the values  $\ifactor(k,x)$ of any algorithm, and therefore a lower bound on the $\ifactor(k)$ of any algorithm. However, to calculate the double integral we need to precisely identify the function~$\alpha$.

\paragraph{Towards finding $\alpha$.}
To calculate $\alpha$, we use what we know from \eqref{eq:N} of how $N$ looks, and rely on the restriction ``on columns''  \eqref{eq:intRestriction}. 
Note that $\mu$ is monotonically decreasing in $x$ and  tending to 0 as $x$ goes to infinity. Therefore, our $N(x,t)$ is non-decreasing in $x$ and at some point reaches 1. 
Mark $\gamma$ the $x$ where $N(x,t)$ becomes~1.  Note that $\gamma$ is a function of $t$ (yet, for readability,  we don't write $\gamma(t)$).
Since $\mu$ is continuous, $\gamma$ is where $\alpha/\mu(\gamma)^{1/(k-1)} = 1$, or in other words, 
$\alpha = \mu(\gamma)^{1/(k-1)}$.
We will find $\alpha$ using our restriction \eqref{eq:intRestriction}:
\BEQ
t = \int_s^\infty(1- N(x,t)) dx = \int_s^{\gamma}  1 - \alpha \mu(x)^{-\inv{k-1}} dx 
= \gamma - s  - \int_s^{\gamma}  \alpha \mu(x)^{-\inv{k-1}} dx 
\EEQ
It follows that:
\BNEQ 
\gamma - t - s & =  
\alpha \int_s^{\gamma} \mu(x)^{-\inv{k-1}} dx \label{eq:basicAlpha} \\
& = \mu(\gamma)^{\inv{k-1}} \int_s^{\gamma} \mu(x)^{-\inv{k-1}} dx
= \int_s^{\gamma} \left(\frac{\mu(\gamma)}{\mu(x)}\right)^{\inv{k-1}} dx  \label{eq:gamma}
\ENEQ
We have in fact found a restriction on $\gamma$, which, if we manage to solve, will also give us $\alpha$.

\vspace{-10pt}
\paragraph{Simplifying the Double Integral.}

We return to the weighted average:
\BNEQ  \label{eq:ifactor}
\int_0^\infty \int_s^\infty \mu(x) N(x,t)^k dx dt 
& = \int_0^\infty \left( \int_s^{\gamma} \mu(x) \alpha^k \mu(x)^{-\frac{k}{k-1}}dx + \int_{\gamma}^\infty \mu(x)dx \right) dt \nonumber \\
& = \int_0^\infty \left( \alpha^{k-1} \int_s^{\gamma} \alpha \mu(x)^{-\inv{k-1}}dx + \int_{\gamma}^\infty \mu(x)dx \right) dt \nonumber \\
& = \int_0^\infty \left( \mu(\gamma) (\gamma - t - s) - M(\gamma) \right) dt 
\ENEQ
For the last equation, we used \eqref{eq:basicAlpha}, the fact that $\alpha^{k-1} = \mu(\gamma)$, and denoted $M(x) = \int \mu(x) dx$ the indefinite integral.
Note that $M(\infty) = 0$ because as we saw in \eqref{eq:muIntegrable}, the integral $\int_0^\infty \mu(x)$ is defined.

\vspace{-10pt}
\paragraph{A Specific Weighted Average.}

Over any specific  weighted average $\omega$, 
the result  we get for \eqref{eq:ifactor} will be a lower bound on the $\ifactor(k)$ for all algorithms.
We will now choose a specific $\omega$ to work with. We show here a proof with ``rounded corners", and leave the exact version to appendix \ref{apx:exactOmega}.

First rounded corner is that we assume $s = 0$.
Second one is that we take  $\omega(x) = \frac{I}{x}$ where $I = 1 / \int_0^\infty \inv{x}$. This makes sense as a normalization, except that  
the integral does not converge. Yet we assume here that it does. Dealing with these  corners requires the manipulation of involved integrals and a careful choice of approximations.
Otherwise it is easy. We have $\mu(x) = \frac{I}{x^2}$. 	
Take \eqref{eq:gamma}:
\BEQ
\gamma - t 
 = \int_0^\gamma \left(\frac{\mu(\gamma)}{\mu(x)}\right)^{\inv{k-1}} dx 
 = \int_0^\gamma \left(\frac{x}{\gamma} \right)^\frac{2}{k-1} dx 
 = \gamma \int_0^{1} x^\frac{2}{k-1} dx 
 = \gamma \frac{k-1}{k+1}
\EEQ
This gives
$\gamma = \frac{k+1}{2} t$, and
plugging this in our \eqref{eq:ifactor}:
\BEQ 
\int_0^\infty \left( \mu(\gamma) (\gamma - t) - M(\gamma) \right) dt 
& = \int_0^\infty \left( \frac{4I}{t^2(k+1)^2} \left(\frac{k-1}{2}t\right) + \frac{2I}{(k+1)t} \right) dt \\
& = \frac{2I}{k+1} \int_0^\infty \left( \frac{k-1}{k+1} + 1 \right) \inv{t} dt
 =\frac{4kI}{(k+1)^2} \int_0^\infty \inv{t} dt
 = \frac{4k}{(k+1)^2} 
\EEQ
Concluding our proof of Theorem \ref{thm:lowerBound}.

\section{Future Directions}
While the questions and the answers in this paper are simple, the techniques used are rather involved. We would love to see simpler proofs. 
On the other hand, it seems that our techniques may be used for other related problems. For example, when the treasure is hidden not by an adversary, but according to a known distribution. Other intriguing questions concern reducing coordination in search problems that involve multiple treasures. 

\paragraph{Acknowledgments.} We thank Stephan Holzer and Lucas Boczkowski for helpful discussions. 

\clearpage

\appendix

\section{Robustness Proof} \label{apx:robust}

\newtheorem*{theorem:robust}{Theorem \ref{thm:robust}}
\begin{theorem:robust}
Denote $c = \speedup(k)$ when all searchers see the correct order of the boxes. Consider the case where they see a different ordering (possibly including some extra boxes), and mark by $\sigma_s(i)$ the index of box $i$ in the eyes of searcher~$s$. 
If for every $i$, $\lim_{i \rightarrow \infty} \sigma_s(i)/i=1$
then the new speed-up is at least~$c$.
\end{theorem:robust}
\BPF
The probability that box $z$ was not checked by time $t$ is
$
N(\sigma_1(z), t) \cdots N(\sigma_k(z), t)
$.
The expected time until we look into box $z$ is therefore:
\BEQ
\sum_{t=0}^\infty N(\sigma_1(z), t) \cdots N(\sigma_k(z), t)
& \leq
\bracks{\sum_{t=0}^\infty N(\sigma_1(z),t)^k}^\inv{k} \cdots \bracks{\sum_{t=0}^\infty N(\sigma_k(z),t)^k}^\inv{k} 
\EEQ
By a generalized form of H\"{o}lder's inequality \cite{Holder}.
This is exactly:
\BEQ
\bracks{\frac{\sigma_1(z)}{\speedup(k, \sigma_1(z))}}^\inv{k} \cdots 
\bracks{\frac{\sigma_k(z)}{\speedup(k, \sigma_k(z))}}^\inv{k}
\EEQ
Fix a small $\epsilon > 0$, and take large enough $X$ so that,
\begin{itemize}
\I For all $x > X/2$, $\speedup(k,x) > c - \epsilon$. 
\I For all $z > X$, for all $i$, $\sigma_i(z) < (1 + \epsilon) z$. 
\end{itemize}
We then get that taking $z > X$, the expected running time until we peek into box $z$ is at most:
\BEQ
\bracks{\frac{(1+\epsilon)z}{c-\epsilon}}^\inv{k} \cdots 
\bracks{\frac{(1+\epsilon)z}{c-\epsilon}}^\inv{k} 
=
\frac{1+\epsilon}{c-\epsilon}z
\EEQ
As $\epsilon$ is arbitrarily small, we get that the speedup is at least $c$.
\EPF

\section{Upper Bound Claim} \label{apx:upper}

\newtheorem*{claim:boxes}{Claim \ref{claim:boxes}}
\begin{claim:boxes}
	For any $\delta > 0$ and $k > 2/\delta$, we have:
	\BEQ
	\limsup_{x\rightarrow\infty}\inv{x} \sum_{t=x}^\infty \bracks{\prod_{i=x}^t  \frac{i}{i + \delta}}^k \leq \inv{\delta k - 1}
	\EEQ
\end{claim:boxes}
\medskip
\BPF
To prove the claim we use basic properties of the $\Gamma$ function. The first is that for natural numbers $\Gamma(n) = (n-1)!$. Another property is that for any real positive number, $\Gamma(z+1) = z\Gamma(z)$.
So:
\BEQ
\Gamma((t+1) + \delta) 
& = (t + \delta) \Gamma(t + \delta) \\
& = (t + \delta) (t-1 + \delta) \Gamma(t-1 + \delta) \\
& = (t + \delta) (t-1 + \delta) \cdots (x + \delta) \Gamma(x + \delta)
\EEQ
So:
\BEQ
(t + \delta) (t-1 + \delta) \cdots (x + \delta)  = \frac{\Gamma(t+1 + \delta)}{\Gamma(x + \delta)}
\EEQ
We therefore get that:
\BEQ
\prod_{i=x}^t  \frac{i}{i + \delta} = 
\frac{\Gamma(t+1) / \Gamma(x)}{\Gamma(t+1 + \delta) / \Gamma(x+\delta)}
= \frac{\Gamma(t+1)}{\Gamma(t+1+\delta)} \cdot \frac{\Gamma(x+\delta)}{\Gamma(x)}
\EEQ
We use a known property of the $\Gamma$ function: 
\BEQ
\lim_{n\rightarrow\infty} \frac{\Gamma(n + \alpha)}{\Gamma(n) n^\alpha} = 1
\EEQ
So for a given $\epsilon > 0$, and for large enough $x$:
\BEQ
\prod_{i=x}^t \frac{i}{i + \delta} < 
(1+\epsilon) \bracks{\frac{x}{t+1}}^\delta
\EEQ
We now look at the sum (again, for large enough $x$):
\BEQ
\inv{x}\sum_{t=x}^\infty \bracks{\prod_{i=x}^t  \frac{i}{i + \delta}}^k < 
\inv{x}(1+\epsilon)^k \sum_{t=x}^\infty \bracks{\frac{x}{t+1}}^{\delta k}
\EEQ
The sum is less than the integral if we sample at the next point as the function is monotonically decreasing in $t$, so this is at most (leaving the $(1+\epsilon)$ aside for a moment):
\BEQ
\inv{x} \int_{x}^\infty \bracks{\frac{x}{t}}^{\delta k} dt = 
\int_{1}^\infty \inv{t^{\delta k}} dt = \inv{-\delta k + 1} \left(\inv{ t^{\delta k - 1}} \right|_1^\infty = \inv{\delta k - 1}
\EEQ
Taking $x$ to infinity, $\epsilon$ goes to 0, and we get our result.
\EPF

\section{Optimal Function Lemma} \label{apx:optimal}	

Assume $k \geq 2$. We want to prove:	
\newtheorem*{lemma:balance}{Lemma \ref{lm:balance}}
\bigskip
\begin{lemma:balance}
Fix $s\geq 0$ and some $M > 0$. For continuous functions $a: [s,\infty) \rightarrow (0,M]$ and $f : [s,\infty) \rightarrow [0,1]$ where $\int_s^\infty a(x) dx = A$ and $\int_s^\infty (1-f(x))dx = T$, 
the minimum of $\int_s^\infty a(x) f(x)^k dx$, ranging over all possible $f$'s is achieved when
$f(x) = \min(1, \alpha / a(x)^{1/(k-1)})$, where $\alpha$ is a function of $a$ and $T$, and independent~of~$x$. 
Also, fixing~$a$, $\alpha$ is a continuous function of $T$.
\end{lemma:balance}
\bigskip
\noindent The proof proceeds gradually, where we prove a version of this lemma first on a domain of size two. We use it to prove the lemma on any finite domain, and from there we prove it on a countable domain. Finally we prove the full lemma as stated above.
\bigskip 	
\begin{lemma} \label{lm:balance2}	
Fix $T$ such that $0\leq T\leq 2$. Given $a_1, a_2 > 0$, 	
The minimal value of $a_1f_1^k + a_2f_2^k$, where $f_1, f_2 \in [0,1]$ and $f_1 + f_2 = T$ is	
achieved when $f_1 = \min(1, \bracks{a_2/a_1}^\inv{k-1} \cdot f_2)$. Furthermore, the pair $(f_1,f_2)$ achieving this minimum is~unique.
\end{lemma}
\medskip
\BPF
Mark $\alpha = a_2/a_1$, and 
$f_2 = T - f_1$.
We would like to minimize:
\BEQ 
f_1^k + \alpha(T-f_1)^k
\EEQ
We take the derivative w.r.t $f_1$:
\BNEQ \label{eq:firstDerivative}
k\left(f_1^{k-1} - \alpha(T-f_1)^{k-1}\right)
\ENEQ 
This is zero exactly when:
\BNEQ \label{eq:minx}
f_1 = \alpha^\inv{k-1} \cdot (T-f_1)
\ENEQ 
So,
\BNEQ \label{eq:positive}
&f_1 =  \frac{\alpha^\inv{k-1}}{1 + \alpha^\inv{k-1}} T
\ENEQ
We take the second derivative (the first was \eqref{eq:firstDerivative}),
\BEQ
k(k-1)\left(f_1^{k-2} + \alpha(T-f_1)^{k-2}\right)
\EEQ 
If we look at $f_1$'s in the range $[0,T]$, this is always positive, meaning our function is U-shaped in this range, since by \eqref{eq:positive} the minimum is somewhere in $[0,T]$. Recall that $f_1\in~[0,1]$. If the bottom of the U is in~$[0,1]$ then as we've seen in \eqref{eq:minx} we get the lemma. 
Otherwise it must be somewhere in $(1,T]$, and so our minimum would be at 1.
\EPF
\medskip
\begin{lemma} \label{lm:finiteBalance}
Fix $T$ such that $0\leq T\leq n$, and fix  $a_1, \dots, a_n > 0$. The minimal value of $\sum_{i=1}^n a_if_i^k$, assuming that $f_1,\ldots,f_n \in [0,1]$ and $\sum_i f_i = T$, is achieved when 
$f_i = \min(1, \alpha / a_i^{1/(k-1)})$,
where $\alpha$ is a function of the $a_i$'s and $T$.
This minimum is unique.
\end{lemma}
\medskip
\BPF
Since the set of solutions $(f_1, ... f_n)$ is a compact space, and our function $\sum_{i=1}^n a_if_i^k$ is continuous, then its image is a compact part of the real line, and so has a minimum.
This means there is an optimal solution.

Take some $i$, such that $1 \neq i \leq n$. We can rebalance the values of $f_i$ and $f_1$ as we wish as long as the sum of $f_i + f_1$ remains the same. If we do so according to Lemma \ref{lm:balance2} it will improve the solution, unless:
\BEQ
f_i = \min(1, \bracks{\frac{a_1}{a_i}}^\inv{k-1} f_1)
\EEQ
Writing $\alpha =  a_1^\inv{k-1} f_1$, we obtain the form $f_i = \min(1, \alpha / a_i^{\inv{k-1}})$, as required. We next show that  the minimum solution is unique and that  $\alpha$ is a function of the $a_i$'s and $T$.

Any minimal solution will look as above, and it must satisfy:
\BEQ
\sum_{i=1}^n \min(1, \alpha / a_i^\inv{k-1}) = T
\EEQ
The left hand side is strictly monotone in $\alpha$, starts from $0$ if $\alpha$ is 0, and is at maximum $n$ if $\alpha$ is large enough. It is also continuous in $\alpha$. Therefore, there is a unique $\alpha$ that solves this, as long as $T \leq n$, and this unique value depends only on the $a_i$'s and on $T$.
\EPF
\bigskip
We will now expand Lemma \ref{lm:finiteBalance} to a countable number of points, and later to the continuous case. It would be great if our compactness argument from Lemma \ref{lm:finiteBalance} would work here but unfortunately, the solution spaces cease to be compact. So we have to work a little harder.
\bigskip

\begin{lemma} \label{lm:countableBalance}
Fix $T$ such that $T \geq 0$. Given $a_1,a_2,\ldots > 0$, where $\sum_i a_i = A$, 
the minimal value of $\sum_i a_i f_i^k $, where 
all $f_i \in [0,1]$ and $\sum_i (1-f_i) = T$, 
is achieved when: 
\BEQ
f_i = \min(1, \alpha / a_i^\inv{k-1})
\EEQ
where $\alpha$ is a function of the $a_i$'s and $T$. 
\end{lemma}
\medskip
\BPF
Taking the $f_i$'s as suggested, {\em i.e.,} $f_i = \min(1, \alpha / a_i^\inv{k-1})$, we first have to show that there exists a unique $\alpha>0$ that satisfies the requirement $\sum_i (1-f_i) = T$. 
Examine this sum as it behaves as a function of~$\alpha$. We next show that for large $\alpha$ the sum is zero, and that as $\alpha$ goes to zero the sum tends to infinity.
To show that for large $\alpha$ the sum is zero, observe first, that the sum of the $a_i$'s converges,
and therefore there is a maximal one: $a_{\max}$. Take $\alpha\geq a_{\max}^{1/(k-1)}$. 
Then, for all~$i$, $f_i = 1$, and $\sum_i (1-f_i) = 0$.
On the other hand, for any $\alpha > 0$, there is a finite number of the $a_i$ that satisfy $a_i^{1/(k-1)} > \alpha$, and only these will have $f_i < 1$, so $\sum_i (1-f_i)$  converges. Furthermore, we can get this sum to be as large as we want by taking $\alpha$ close to~$0$.  
Note that the sum is continuous as a function of $\alpha$, and that for $\alpha$ such that $0<\alpha\leq a_{\max}$, 
it is strictly decreasing. Therefore,   
there is exactly one $\alpha$ that satisfies the requirement $\sum_i (1-f_i) = T$.

Our next goal is to prove that no other $g = (g_1,g_2,\dots)$ satisfies the aforementioned requirements and achieves a smaller value than our suggested solution $f = (f_1,f_2,\dots)$. Assume by way of contradiction, that there is such $g = (g_1,g_2,\dots)$   improving our suggestion by some $\delta>0$, that is,
$\sum_i a_i g_i^k \leq \sum_i a_i f_i^k -\delta$.

For a small $\epsilon > 0$ to be determined, 
take large enough $n$ so that all  of the following are satisfied:
\begin{itemize}
\I $\sum_{i=n+1}^\infty a_i < \epsilon$~,
\I $\sum_{i=n+1}^\infty (1-f_i) < \epsilon$~,
\I $\sum_{i=n+1}^\infty (1-g_i) < \epsilon$~,
\I $\sum_{i=n+1}^\infty a_i f_i^k < \frac{\delta}{2}$~.
\end{itemize}
\medskip

\noindent From this we get:
\BNEQ  \label{eq:targetIneq} 
\sum_{i=0}^n a_i f_i^k - \sum_{i=0}^n a_i g_i^k 
& =
\left(\sum_{i=0}^\infty a_i f_i^k - \sum_{i=0}^\infty a_i g_i^k \right) -
\left(\sum_{i=n+1}^\infty a_i f_i^k - \sum_{i=n+1}^\infty a_i g_i^k \right) \nonumber \\
&> \delta - \left(\frac{\delta}{2} - 0\right) =  \frac{\delta}{2}
\ENEQ
We can assume that for $i=1,2,\ldots,n$, we have $g_i = \min(1, \beta/ a_i^\inv{k-1})$, while setting $\beta$ to keep $\sum_{i=1}^n g_i$ unchanged. Indeed, by Lemma \ref{lm:finiteBalance} this can only decrease the sum $\sum_{i=0}^n a_i g_i^k$ and will therefore still satisfy the above inequality. 

We also have:
\BEQ
\left|\sum_{i=0}^n f_i - \sum_{i=0}^n g_i \right|  
&= 
\left|\sum_{i=0}^n (1-f_i) - \sum_{i=0}^n (1-g_i) \right|  \\
&= 
\left|\left(\sum_{i=0}^\infty (1-f_i) - \sum_{i=0}^\infty (1-g_i)\right)  - 
\left(\sum_{i=n+1}^\infty (1-f_i) - \sum_{i=n+1}^\infty (1-g_i)\right) 
\right| \\
&= 
\left| \sum_{i=n+1}^\infty (1-f_i) - \sum_{i=n+1}^\infty (1-g_i) \right| \\
& \leq
\left|\sum_{i=n+1}^\infty (1-f_i)\right| + \left|\sum_{i=n+1}^\infty (1-g_i) \right|
< 2\epsilon \\
\EEQ
On the other hand, note that for all $i$ such that $1 \leq i \leq n$, we have
$f_i=\min(1, \alpha / a_i^{1/(k-1)})$ and $g_i = \min(1, \beta/ a_i^{1/(k-1)})$. It follows that either all $f_i \geq g_i$ or the other way around (depending on whether $\alpha > \beta$ or vice versa).
Hence:

\BEQ
\sum_{i=0}^n |f_i - g_i|= \left|\sum_{i=0}^n f_i - \sum_{i=0}^n g_i \right|  
 < 2\epsilon \\
\EEQ
Next, let us look at the following difference between the partial sums:
\BEQ
\sum_{i=0}^n a_i f_i^k - \sum_{i=0}^n a_i g_i^k
&= \sum_{i=0}^n a_i (f_i^k - g_i^k) \\ 
&= \sum_{i=0}^n a_i (f_i - g_i)(f_i^{k-1} + f_i^{k-2}g_i + \dots + g_i^{k-1}) \\
&\leq k \cdot a_{max} \sum_{i=0}^n  |f_i - g_i| \\
&\leq  k \cdot a_{max} \cdot 2 \epsilon \\
\EEQ
Taking small enough $\epsilon$, we obtain:
\BEQ
\sum_{i=0}^n a_i f_i^k - \sum_{i=0}^n a_i g_i^k < \frac{\delta}{2}
\EEQ
Contradicting \eqref{eq:targetIneq}. This completes the proof of the lemma.
\EPF

\bigskip
Finally, we arrive at the proof of Lemma \ref{lm:balance}.	
\BPF
First we have to prove that there is an $\alpha$, such that $\int_s^\infty (1-f(x))dx = T$, where $f = \min(1, \alpha / a(x)^\inv{k-1})$ is the proposed optimal function. Define $I = \stset{x}{a(x)^\inv{k-1} \geq \alpha}$.
We look at the integral as a function of $\alpha > 0$:
\BEQ
\int_s^\infty (1-f(x))dx &= \int_s^\infty \bracks{1 - \min(1, \alpha/a(x)^\inv{k-1})}dx \\
&= \int_I \bracks{1 - \alpha/a(x)^\inv{k-1}} dx \leq |I|
\EEQ
We have:
\begin{itemize}
\I 
Anywhere within $I$, $a(x) > const$, and so $I$ must be of finite measure, otherwise $\int_s^\infty a(x) dx$ does not converge. So this integral is always defined. 
\I
The integral is strictly decreasing in $\alpha$ , since increasing $\alpha$ shrinks $I$ (and as $a$ is continuous will make it strictly smaller) and increases $f(x)$.
\I
Since for all $x$, $a(x) \leq M$, taking $\alpha \geq M^\inv{k-1}$ we get that the integral is 0.
\I
If we take $\alpha$ towards 0, $I$ increases its size and we can make it as large as we wish. If we want the integral to be larger than some $M$, then we take $\alpha$ small enough to make $|I| > 2M$, and now take half that $\alpha$. The new $I$ contains the previous one, and all the $x$'s that were in the old $I$ now have $f(x) \leq \inv{2}$ so the integral is at least $M$. 
This means that as $\alpha$ goes to 0, the integral goes to infinity.  
\I
The integral is a continuous function of $\alpha$.
\end{itemize}
Putting these together, we see that there is exactly one $\alpha$ that fits.
Note also, that if we think of $\alpha$ as a function of $T$, then it is also continuous, since it is the inverse function of a continuous strictly monotone function.

Assume there is some other function $g$ satisfying the requirements, that improves on the target function by $\delta$.
Take small enough $d$, so that taking points $x_1=s+d, x_2=s+2d, ...$, all of the following are satisfied (marking $a_i = a(x_i), f_i = f(x_i)$ and $g_i = g(x_i)$), where $\epsilon > 0$ will be determined later:
\BE[label=(\alph*)]
\I \label{item:1}
$\left| \int_s^\infty a(x) f(x)^\inv{k-1} dx - d\sum_i a_i f_i^\inv{k-1} \right| < \epsilon$
\I \label{item:2}
$\left| \int_s^\infty a(x) g(x)^\inv{k-1} dx - d\sum_i a_i g_i^\inv{k-1} \right| < \epsilon$
\I \label{item:4}
$\left| T - d\sum_i \bracks{1 - f_i} \right| < \epsilon$
\I \label{item:5}
$\left| T - d\sum_i \bracks{1 - g_i} \right| < \epsilon$
\EE
From \ref{item:1} and \ref{item:2}:
\BNEQ \label{eq:continuous1}
d\sum_i a_i f_i^\inv{k-1} - 
d\sum_i a_i g_i^\inv{k-1} 
> \delta - 2\epsilon 
\ENEQ
and from \ref{item:4} and \ref{item:5}:
\BEQ
\left|d\sum_i (f_i - g_i) \right| < 2\epsilon
\EEQ
We proceed:
\BEQ
d\sum_i a_i f_i^k - d\sum_i a_i g_i^k
&= d\sum_i a_i (f_i - g_i) (f_i^{k-1} + f_i^{k-2}g_i + \dots + g_i^{k-1}) \\
&\leq k d M \sum_i  |f_i - g_i| \\
\EEQ
By Lemma \ref{lm:countableBalance}, the $f_i$'s give minimal $\sum_i a_i f_i^k$ amongst all such series that have the same $\sum_i (1 - f_i)$.
We can assume that the $g_i$'s are of the same form, since by lemma \ref{lm:countableBalance} changing them to this form while keeping $\sum_i (1-g_i)$ will only improve their value, and so will keep \eqref{eq:continuous1} valid. We therefore know that either for all~$i$, $f_i \geq g_i$ or for all~$i$, $g_i \geq f_i$.  
So:
\BEQ
d\sum_i a_i f_i^k - d\sum_i a_i g_i^k
\leq k M \cdot \left|d\sum_i (f_i - g_i) \right|
< k M \cdot 2 \epsilon \\
\EEQ
 Taking a small enough $\epsilon$, this will be smaller than 
$\delta - 2\epsilon$, contradicting \eqref{eq:continuous1}. 
\EPF

\section{Exact Choice of $\omega$} \label{apx:exactOmega}

Recall our situation. What we know of $\gamma$ is:
\BNEQ \label{eq:apxGamma}
\gamma - t - s &  
= \int_s^{\gamma} \left(\frac{\mu(\gamma)}{\mu(x)}\right)^{\inv{k-1}} dx
\ENEQ
And the integral we wish to calculate is:
\BNEQ \label{eq:apxIntegral}
\int_0^\infty \left( \mu(\gamma) (\gamma - t - s) - M(\gamma) \right) dt 
\ENEQ
Every weighted average $\omega$ we take will give us a $\gamma$ from \eqref{eq:apxGamma}, and from that we can calculate the value of \eqref{eq:apxIntegral} which will be a lower bound on $\ifactor(k)$.
The $\omega$ we take is $\omega(x) = \frac{I}{x^{a-1}}$ for some $a > 2$, where,
\BNEQ \label{eq:I}
I = 1 / \int_s^\infty \inv{x^{a-1}} = -1 / \left( \inv{(a-2)x^{a-2}} \right|_s^\infty 
= (a-2) s^{a-2} 
\ENEQ
This works fine as long as $a>2$.
We then get:
\BEQ
\mu(x) = \frac{I}{x^a} \mbox{~~~~~~and~~~~~~}
M(x) = -\frac{I}{(a-1) x^{a-1}}
\EEQ
This $\mu$ satisfies our requirements: it is strictly positive, bounded, continuous and monotonically decreasing.  
We find $\gamma$ from \eqref{eq:apxGamma}:
\BEQ
\gamma - t -s 
& = \int_s^\gamma \left(\frac{\mu(\gamma)}{\mu(x)}\right)^{\inv{k-1}} dx 
= \int_s^{\gamma} \left(\frac{x}{\gamma} \right)^\frac{a}{k-1} dx 
= \gamma \int_{\frac{s}{\gamma}}^{1} x^\frac{a}{k-1} dx \\
& = \gamma \inv{1+\frac{a}{k-1}} \left( x^{1 + \frac{a}{k-1}} \right|_{\frac{s}{\gamma}}^1 
= \gamma \frac{k-1}{a + k - 1} \left( 1 - \bracks{\frac{s}{\gamma}}^\frac{a+k-1}{k-1} \right)
\EEQ
Since $\gamma \geq s$, and $\frac{a+k-1}{k-1} > 1$, then
$\bracks{\frac{s}{\gamma}}^\frac{a+k-1}{k-1} < \frac{s}{\gamma}$,
and therefore:
\BEQ
\gamma \frac{k-1}{a+k-1} > \gamma-t-s > \gamma \frac{k-1}{a+k-1} (1- \frac{s}{\gamma})
\EEQ
Left side gives:
\BEQ
\gamma \frac{a}{a+k-1} < t + s \\
\gamma  < \frac{a+k-1}{a} (t + s) \\
\EEQ
Right side gives:
\BEQ
\gamma-t-s  > \frac{k-1}{a+k-1} (\gamma - s) \\
(\gamma - s) \frac{a}{a+k-1} > t \\
\gamma > \frac{a+k-1}{a}t + s \\
\EEQ
Plugging this in our formula \eqref{eq:apxIntegral}:
\BEQ 
\int_0^\infty \left( \mu(\gamma) (\gamma- t-s) - M(\gamma) \right) dt 
& = I \int_0^\infty \left(\inv{\gamma^a} (\gamma-t -s) + \inv{a-1} \cdot \inv{\gamma^a} \gamma \right) dt\\
& = I \int_0^\infty \inv{\gamma^a} \left( \gamma - t -s + \inv{a-1} \gamma \right) dt\\
& = I \int_0^\infty \inv{\gamma^a} \left( \frac{a}{a-1}\gamma - t - s\right) dt \\
& \geq I \int_0^\infty \inv{\gamma^a} \left( \frac{a}{a-1} \bracks{\frac{a+k-1}{a}t +s} - t - s \right) dt \\
& = I \int_0^\infty \inv{\gamma^a} \left( \frac{k}{a-1}t + \inv{a-1}s \right)dt \\
& = I \frac{k}{a-1}\int_0^\infty \frac{t + \frac{s}{k}}{\gamma^a} dt \\
& \geq I \frac{k}{a-1}\int_0^\infty \frac{t + \frac{s}{k}}{\bracks{\frac{a+k-1}{a}(t + s)}^a} dt \\
& = I \frac{k}{a-1} \bracks{\frac{a}{a+k-1}}^a\int_0^\infty \frac{t + \frac{s}{k}}{(t + s)^a} dt
\EEQ
Let's just look at the integral. By changing $t$ to $t-s$:
\BEQ
\int_s^\infty \frac{t - \frac{k-1}{k}s}{t^a} dt 
&= \int_s^\infty \inv{t^{a-1}} dt  - \frac{k-1}{k}s \int_s^\infty \inv{t^a} dt \\
&= 1/I - \frac{k-1}{k}s \left(-\inv{(a-1)t^{a-1}}\right|_s^\infty \\
&= 1/I - \frac{k-1}{k}s \inv{(a-1)s^{a-1}} \\
&= 1/I - \frac{k-1}{k(a-1)s^{a-2}}
\EEQ
Plugging this back in:
\BEQ
I \frac{k}{a-1} \bracks{\frac{a}{a+k-1}}^a
\left( 1/I - \frac{k-1}{k(a-1)s^{a-2}} \right) =
\frac{k}{a-1} \bracks{\frac{a}{a+k-1}}^a
- T \frac{I}{s^{a-2}}
\EEQ
Where $T$ is a constant related to $k$ and $a$ that remains bounded as $a$ approaches 2.
Plug in $I$ from \eqref{eq:I} and the second part becomes:
\BEQ
T \frac{(a-2)s^{a-2}}{s^{a-2}} = T(a-2) 
\EEQ
So the whole sum is:
\BEQ
\frac{k}{a-1} \bracks{\frac{a}{a+k-1}}^a - T(a-2)
\EEQ
Taking $a$ towards $2$, we can get as close as we wish to:
\BEQ
\frac{k}{2-1} \left(\frac{2}{2+k-1}\right)^2 =
\frac{4k}{(k+1)^2} 
\EEQ
Concluding our proof.


\clearpage



\begin{thebibliography}{10}

\bibitem{AlonZvika}
N. Alon, C. Avin, M. Kouck, G. Kozma, Z. Lotker, and M. R. Tuttle.
\newblock  Many Random Walks Are Faster Than One.
\newblock {\em Combinatorics, Probability \& Computing}, 20(4), 481--502, 2011.


\bibitem{BCR91}
R. A. Baeza-Yates, J. C. Culberson and G.J.E. Rawlins.
\newblock Searching in The Plane.
\newblock {\em Information and Computation}, (106), (2), 234--252, 1991.


\bibitem{CFR09}
C. Cooper, A. M. Frieze, and T. Radzik.
\newblock Multiple Random Walks in Random Regular Graphs.
\newblock  {\em SIAM J. Discrete Math}, (2009), 23(4), 1738-1761.


\bibitem{DFG06}
E.~D. Demaine, S.~P. Fekete, and S.~Gal.
\newblock Online searching with turn cost.
\newblock {\em Theoret. Comput. Sci.}, 361(2-3), 342--355, 2006.


\bibitem{ES11}
R. Elsasser and T. Sauerwald.
\newblock  Tight bounds for the cover time of multiple random walks.
\newblock {\em Theor. Comput. Sci. 412}, 24, 2623--2641, 2011.



\bibitem{Tonelli}
E. DiBenedetto.
\newblock Real analysis.
\newblock {\em Birkhäuser Advanced Texts: Basler Lehrbücher}, Boston, MA: Birkhäuser Boston, Inc.


\bibitem{Emek-ants}
Y. Emek, T. Langner, D. Stolz, J. Uitto, and R. Wattenhofer. 
\newblock How Many Ants Does It Take To Find the Food? 
\newblock {\em SIROCCO}, pages 263-278, 2014.

\bibitem{Uitto}
Y. Emek, T. Langner, J. Uitto, and R. Wattenhofer. 
\newblock Solving the ANTS problem with Asynchronous Finite State Machines. 
\newblock {\em ICALP} 2014.


\bibitem{FK12}
O. Feinerman and A. Korman.
\newblock  Memory Lower Bounds for Randomized Collaborative Search and Implications to Biology.
\newblock Proc. 26th  International Symposium on Distributed Computing (DISC), 61--75,  2012.



\bibitem{FKLS}
O. Feinerman, A. Korman, Z. Lotker and J.S. Sereni.
\newblock  Collaborative Search on the Plane without Communication.
\newblock Proc. 31st ACM Symp. on Principles of Distributed Computing (PODC),77--86,  2012.


\bibitem{Holder}
H. Finner. 
\newblock A Generalization of Holder's Inequality and Some Probability Inequalities. 
\newblock {\em Ann. Probab.} 20 (1992), no. 4, 1893--1901.


\bibitem{JaJa}
J. J\'aJ\'a. 
\newblock Introduction to Parallel Algorithms.
\newblock {\em  Addison-Wesley}, 1992. 

\bibitem{KRT96}
M. Kao, J. H. Reif, and S. R. Tate.
\newblock  Searching in an Unknown Environment: An Optimal Randomized Algorithm for the Cow-Path Problem.
\newblock  {\em Journal of Inf. Comput.}, 63--79, 1996.

\bibitem{KSW86}
R. M. Karp, M. Saks, and A. Wigderson.
\newblock On a search problem related to branch-and-bound procedures.
\newblock {\em In Proc. of the 27th Ann. Symposium on Foundations of Computer Science (FOCS)}, 19--28, 1986.


\bibitem{Tobias}
T. Langner, D. Stolz, J. Uitto and R. Wattenhofer.
\newblock Fault-Tolerant ANTS.
\newblock In {\em DISC} 2014.



\bibitem{Tom}
F.T. Leighton.
\newblock Introduction to Parallel Algorithms and Architectures: Arrays, Trees, Hypercubes. 
\newblock {\em Morgan Kaufmann Publishers Inc.}, San Francisco, CA, USA, 1992. 

\bibitem{Nancy-PODC}
C. Lenzen, N.A. Lynch, C. Newport, and T. Radeva. 
\newblock Trade-offs between Selection Complexity and Performance when Searching the Plane without Communication. 
\newblock {\em PODC} 2014.

\bibitem{BOINC}
https://boinc.berkeley.edu

\end{thebibliography}
\end{document}